# Does social media information affect individual investor disposition effect? Evidence from Xueqiu[1]

Siliu Chen, Fei Ren*

School of Business, East China University of Science and Technology, Shanghai, China

* fren@ecust.edu.cn (FR)

## Abstract

The irrational behavior of investors selling profitable assets too early while holding onto losing assets for too long is known as the disposition effect. Due to the development of the Internet, the information environment for individual investors has been greatly improved. As an important source of information for individual investors, whether social media can improve investors' behavioral biases and return to rational expectations is a question worth studying. Based on the post data and actual trading data of the social investment platform Xueqiu.com, this paper studies the impact of social media information on the disposition effect of individual investors. The research results show that social media information can significantly reduce the disposition effect. Furthermore, it is through negative information that social media information reduces the disposition effect. When presented with negative information, individual investors will gradually become more rational in adjusting their positions. At the individual level, factors such as investment experience, users followed, region, and gender can all influence the effectiveness of the information acquired by individual investors in reducing the disposition effect.

[1] Published in PLOS ONE. DOI: https://doi.org/10.1371/journal.pone.0328547

# 1. Introduction

The disposition effect refers to a robust trading phenomenon in which investors tend to hold losers too long and sell winners too soon. This financial anomaly, which goes against the rational person assumption and expected utility theory of traditional finance, was first proposed and named by Shefrin and Statman [1] in 1985. Subsequent studies have shown that the disposition effect is widespread in global financial markets such as stocks [2-4], funds [5, 6], and futures [7]. Odean [3] first proposed the method of comparing the difference in the proportion of realized gains and losses to measure the disposition effect, and demonstrated its existence in the U.S. stock market. Lu et al. [4] used a large sample of account data from Chinese stock market investors covering 2011 to 2017 to study the disposition effect. They found that the Chinese stock market exhibits a disposition effect, which becomes more pronounced with shorter holding periods. Additionally, the study by Lu et al. [4] also found that compared to American investors, Chinese investors exhibit a preference for recouping losses and show a greater disparity in sensitivity to both gains and losses. Furthermore, some scholars have studied the relationship between investor characteristics and the disposition effect. Brown et al. [8] found that institutional investors, individual investors, and foreign investors participating in the Australian stock market all exhibited a certain degree of disposition effect. However, individual investors and foreign investors showed a higher disposition effect compared to institutional investors with professional expertise. Moreover, the disposition effect of individual investors varies depending on their characteristics. Studies on individual investors have found that experienced investors are more accurate in predicting the stock market and exhibit a lower disposition effect [9]; investors in first-tier cities have a stronger disposition effect than those in other regions [5]; and female investors, who generally have lower levels of financial knowledge[10], higher levels of risk aversion[11], and greater susceptibility to regret [12] compared to male investors, demonstrate a stronger disposition effect [5, 9, 12].

Given the established fact that individual investors exhibit a disposition effect, this study examines whether information acquisition by investors can reduce this effect. With the development of the Internet, channels for Chinese investors to obtain information have changed. Various websites and social media, using the Internet as a medium for dissemination, have taken a dominant position, while the roles of traditional media such as television, radio, newspapers, and magazines have gradually weakened [13]. Unlike traditional media, social media allows users to freely express their opinions on stocks and share investment experiences anytime and anywhere, enabling other users to engage in discussions and exchange information. Due to the timely and low-cost exchange and dissemination of information, information on social media has exploded, significantly improving the information environment for investors. Jin and Li [14] studied the value discovery function of social media and found that stock forums contain undiscovered company-specific information with investment value. Jiang and Yan [15] argue that forum information interacts through three methods: information acquisition and absorption, information exchange and correction, and cognitive formation and expectation, thereby influencing investors' decisions. Therefore, this paper argues that social media information can enhance the information environment for investors, helping them to form expectations about future stock returns. This enables investors to base their investment decisions on future asset returns rather than past gains and losses, thereby reducing the behavioral bias of selling profitable assets while holding onto losing ones. However, there is currently no literature studying disposition effects from the perspective of investor information acquisition. Studying the impact of social media information on disposition effect would enhance our understanding of individual investor trading behaviors in the context of the internet era, which is crucial for future theoretical modeling.

We use post data and actual trading data from Xueqiu.com, China's leading investor social platform, to study the impact of social media information on the disposition effect of individual investors for the first time, and reveals its underlying influence mechanism. The results show that social media information can significantly reduce the disposition effect of individual investors. Furthermore, it is through negative information that social media information reduces the disposition effect. When individual investors absorb negative information, they gradually become more rational in adjusting their positions. At the individual level, factors such as investment experience, users followed, region, and gender can all affect the effectiveness of the information acquired by individual investors in reducing disposition effect.

This paper contributes to the literature on the disposition effect in the following aspects: First, using data from the social investment platform Xueqiu.com, this study is the first to verify that social media information can reduce the disposition effect of individual investors. This finding highlights the important role of social media information in the study of disposition effect and enriches the relevant theoretical research. Second, this paper clarifies the underlying mechanism by which social media information reduces the disposition effect of individual investors, finding that the impact of positive and negative information on the disposition effect follows different paths, and that negative information leads to more rational trading behavior by reducing the tendency to sell winners and hold losers. Third, in addition to existing research indicating that the disposition effect varies with individual investor characteristics, we find that individual investor characteristics also influence investors' ability to process social media information, thereby affecting the disposition effect. This finding extends the research on investor characteristics and the disposition effect.

# 2. Literature Review and Research Hypothesis

## 2.1. Disposition Effect

Shefrin and Statman [1] were the first to discover and propose the concept of the disposition effect, which describes investors' tendency to hold onto losing assets too long and sell winning assets too soon. Since then, numerous scholars both at home and abroad have studied the disposition effect. Odean [3] studied data from 10,000 accounts provided by a nationwide discount brokerage in the U.S. from 1987 to 1993 and proposed the classic disposition effect measure, PGR-PLR. He found that the proportion of profitable stocks sold was higher than that of losing stocks, demonstrating the existence of the disposition effect in the U.S. stock market. Zhao and Wang [16] used data from nearly 10,000 accounts provided by a brokerage firm to be the first to identify the existence of the disposition effect in the Chinese stock market. Xiao et al. [9] analyzed transaction data from 30,000 accounts provided by a brokerage firm, using the classic PGR-PLR measure and the Cox proportional hazard model to demonstrate the existence of the disposition effect in the Chinese securities margin market. Furthermore, they investigated the factors influencing the disposition effect through survival analysis and found that gender, age, and investment level all affect investors' disposition effect. Specifically, female investors exhibit a stronger disposition effect compared to male investors. Based on age groups, middle-aged investors have the strongest disposition effect, followed by young investors, while elderly investors display the lowest disposition effect. Additionally, investors with higher investment levels tend to have a lower disposition effect. Cao et al. [12] used experimental methods to examine the gender differences of disposition effect and its influence mechanism. They found that stock trading decisions of female investors are more likely to be affected by regret and deviate from rationality, resulting in a significantly higher disposition effect compared to male investors. Da Costa et al. [17] developed a computer program to simulate the stock market, using both experienced and inexperienced investors as subjects. They found that both groups exhibited the disposition effect, but the more experienced investors were less affected. Da Silva and Mendes [18] analyzed transaction-level data from 35,508 individuals at a major Portuguese financial institution between 1998 and 2017, finding that while education and experience can reduce the disposition effect, neither fully offsets it. Wu et al. [5] used data from over 400,000 investor accounts across five stock open-end funds in a Chinese mutual fund company and employed survival analysis to test fund investors' disposition effects. They found that Chinese fund investors exhibited a significant disposition effect overall. Specifically, male investors showed weaker disposition effects compared with female investors; younger investors exhibited weaker disposition effects compared with middle-aged and older investors; investors who opened accounts at securities companies had weaker disposition effects compared with those who opened accounts at banks; and investors in first-tier cities demonstrated stronger disposition effects compared with those in other regions. Unlike the above-mentioned literature, which studies the heterogeneity of individual investors based on socio-demographic characteristics, Hwang et al. [19] identify eight distinct trading patterns among individuals using the machine learning technique of deep clustering. They reveal that some clusters exhibit unique trading behaviors and distinct distributions of realized profits compared to others. Furthermore, some researchers have delved into the forms of the disposition effect. Ben-David and Hirshleifer [20] randomly selected account data from 10,000 individual investors in the U.S. stock market and found that the probability of selling exhibited a V-shaped asymmetric pattern: it increased with the level of gains (or losses), with a faster increase in the probability of selling profitable stocks compared to losing ones. This demonstrates the disposition effect, as evidenced by the higher probability of selling profitable stocks relative to losing ones. Lu et al. [4] analyzed a large sample of account data from Chinese stock market investors provided by brokers from 2011 to 2017. They found that Chinese individual investors also exhibit an asymmetric V-shaped disposition effect. However, there are significant differences compared with the V-shaped disposition effect observed in U.S. investors. Further research revealed that when the holding period return of Chinese investors shifted from negative to positive, the probability of selling stocks increased sharply, a phenomenon not observed in U.S. investors. Additionally, the sensitivity of selling behavior to gains and losses is greater in Chinese investors, showing a more pronounced asymmetry in the disposition effect. Additionally, Ulku et al. [21] studied the aggregate trading activities of different types of investors in the stock market across multiple countries during the COVID-19 pandemic. They found that individual investors' buying was driven by a sequence of their usual contrarian trading behavior followed by a unique positive shock to retail investor demand for equities, which made them winners in the worldwide negative bubble. The study also revealed that, compared to margin traders, cash traders exhibited more contrarian trading and signs of the disposition effect. Welch [22] studied the trading behavior of users on Robinhood, a platform that attracts tiny investors. The study discovered that these investors tended to buy stocks with high past trading volume and stocks of firms with products that they were familiar with. Furthermore, they increased their holdings in individual stocks when stock prices increased or decreased greatly, and their timing and steadfastness contributed to their good portfolio returns. An et al. [23] explored the relationship between portfolio return and the disposition effect, finding that the disposition effect for a stock significantly weakens if the portfolio is at a gain, but is large when it is at a loss.

## 2.2. Social Media Information and Disposition Effect

Due to the development of social media, information can be exchanged and disseminated quickly and at low cost, leading to an explosive growth and improving the information environment for investors. On the one hand, social media provide valuable information that reduces information asymmetry. Xiong et al. [24] studied the relationship between the stock forum and the stock market using data from Eastern Money.com from January 2011 to June 2014. Their results showed a bidirectional predictive association between stock forum sentiment and stock returns, message volume and trading volume, as well as disagreement and volatility, indicating that the stock forum contained information not yet reflected in current stock prices. Zheng et al. [25] studied the effect of internet information communication on stock market herding using A-share return data and posting volume data from the stock forum of Eastmoney.com from January 2009 to August 2013. They found that internet information communication could reduce herding in the stock market, inhibit the continued spread of herding behavior, and enhance market efficiency. Zheng et al. [26] study the relationship between information interactions among investors and market price efficiency using a sample of Chinese A-share listed companies. They find that the intensity of investor information interaction is significantly negatively associated with stock price synchronization. The mechanism analysis suggests that social media facilitates not only information transmission but also interactions among investors, improving their information acquisition and processing abilities. This, in turn, promotes the integration of idiosyncratic information into stock prices, thereby reducing stock price synchronization. On the other hand, social media is full of noise. Xiao et al. [27] classified posts in the Eastmoney.com stock forum that lack firm-level trait information as noise. They examined the relationship between this noise and stock price synchronization and found that such noise induces irrational investment decisions, thereby reducing the synchronization of stock price changes. Further, scholars have investigated the influence of social interaction on the disposition effect. Based on online social and simulated trading data from the social investment platform Xueqiu.com, Sun et al. [28] analyzed the impact of social interactions on the investor disposition effect from four dimensions: social centrality, social release content, social engagement, and social attention. They found that higher social engagement, particularly during trading hours, reduces the disposition effect by increasing the proportion of realized losses. Gemayel and Preda [29] studied the disposition effect of investors in different environments using social platform trading data and traditional trading data, respectively. They found that investors in environments where trading behaviors can be observed by others tend to be more cautious in their investment decisions and are more likely to sell losing assets early to reduce book losses. Thus, the disposition effect is lower in environments where trading behavior is observable. Using a unique data source, Heimer [30] examines the trading behavior of investors before and after they enter social platforms and finds that access to the social network nearly doubles the magnitude of investors' disposition effect. However, there is no existing literature that examines how the information investors obtain from social media influences their disposition effects. By integrating the dual aspects of social media information discussed above, this paper proposes the following research hypotheses:

Hypothesis 1: Social media information can reduce investors' disposition effect.

Hypothesis 2: Social media information can increase investors' disposition effect.

# 3. Research Design

## 3.1. Sample and Data

Xueqiu.com (https://xueqiu.com/) is a social network for investors, officially launched on November 11, 2011. It has since become one of the most active social investment platforms in China. On Xueqiu, users can freely share their opinions on stocks, follow users of interest, and be followed by others. When an investor opens Xueqiu.com, posts from the users they follow appear on their homepage, making it easy to access information from these users. Besides posting, Xueqiu users can create both real and simulated portfolios that can be viewed by others, and they can also review the adjustments made by other users to their portfolios. Since real portfolios are more authentic compared with simulated ones, this study utilizes a Python-based web crawler to gather basic information and real trading data from users on Xueqiu, as well as post data from users whom these real users follow. Ultimately, data related to 11,661 real trading users was collected. The basic information includes user ID, followers, fans, followed stocks, number of posts, region, gender, etc. The trading data includes transaction time, traded stocks, stock position before the transaction, and stock position after the transaction. The post data from the users followed includes the posting time and content. Among the data obtained, the earliest transaction occurred on June 27, 2016, and data collection was completed as of May 31, 2023. Therefore, the sample period is defined as June 27, 2016, to May 31, 2023. After obtaining the raw data, the following cleaning steps were performed on the trading data: (1) Since Xueqiu does not provide complete trading records for real accounts, only the latest 200 trades per user are available; therefore, trades with unavailable purchase prices were excluded; (2) Trades involving new stock subscriptions and their subsequent sales were excluded; (3) Combinations containing non-A-shares were excluded; (4) Combinations with only buying actions and no selling actions were excluded; (5) Combinations with fewer than 3 trades were excluded; (6) Combinations with obvious errors in stock positions were excluded; (7) Combinations where basic information of real account users could not be obtained were excluded. After these exclusions, the sample consists of 129,865 trades from 4,210 real portfolios. Additionally, the required stock closing prices, highest prices, lowest prices, and CSI 300 index closing prices were obtained from Wind and Resset.

## 3.2. Variables' Measurements

### 3.2.1. Dependent Variable: Disposition Effect

This paper uses Odean's [3] method to measure the disposition effect by analyzing the difference between realized gains and losses. According to Odean's [3] research, for each investor on a trading day with sell transactions, the stocks in the account are categorized into four cases: For stocks traded during the period, if the selling price is higher than the cost price, it is defined as a Realized Gain; otherwise, it is defined as a Realized Loss. If there are multiple purchases of the stock, the weighted average price is calculated based on the purchase prices and transaction volumes, serving as the purchase price of the stock. For other stocks held during the period, if the purchase price of the stock is lower than the lowest price of the day, it is defined as a Paper Gain; if the purchase price is higher than the highest price of the day, it is defined as a Paper Loss. All prices are adjusted for stock dividends. The Proportion of Gains Realized (*PGR*) is the ratio of the number of profitable stocks sold to the total number of profitable stocks, and the Proportion of Losses Realized (*PLR*) is the ratio of the number of losing stocks sold to the total number of losing stocks. The calculation method is as follows:

$$Proportion\ of\ Gains\ Realized\ \ (PGR) = \frac{Realized\ Gains}{Realized\ Gains + Paper\ Gains}$$

$$Proportion\ of\ Losses\ Realized\ \ (PLR) = \frac{Realized\ Losses}{Realized\ Losses + Paper\ Losses}$$

*Disposition Effect* = *Proportion of Gains Realized* (*PGR*) - *Proportion of Losses Realized* (*PLR*)

When *PGR* is significantly greater than *PLR*, resulting in $PGR - PLR > 0$, the disposition effect is positive. This indicates that investors tend to sell profitable stocks too early while continuing to hold onto losing stocks, thereby exhibiting the disposition effect.

### 3.2.2. Independent Variables

Posts from users followed by individual investors appear on their homepage, making this information directly accessible to them. This study uses these posts to represent the information available to individual investors. Since forum posts are updated frequently and investors typically focus on the most recent ones, we use a 3-day time window, as proposed by Sun et al. [28]. We measure the logarithm of the number of posts made by these users in the 3 days prior to the investors' trades as a proxy for the social media information they acquire.

### 3.2.3. Control Variables

At the individual level, this study follows Frino et al. [31] by using trading frequency to reflect investors' overconfidence, the number of stocks held to represent portfolio diversification, and the average purchase price of holdings to indicate the nominal price range of stocks within their portfolio. At the market level, market returns are used as a control variable. Detailed variable descriptions can be found in Table 1.

**Table 1.** Variable Definitions.

| Variable Name | Variable Symbol | Meaning |
|---|---|---|
| Proportion of Gains Realized | *PGR* | On trading days with sell transactions, the proportion of profitable stocks sold relative to the total number of profitable stocks |
| Proportion of Losses Realized | *PLR* | On trading days with sell transactions, the proportion of loss-making stocks sold relative to the total number of loss-making stocks |
| Disposition Effect | *DE* | Proportion of Gains Realized - Proportion of Losses Realized |
| Realized Gains | *RG* | The number of profitable stocks among the stocks sold on trading days with sell transactions |
| Realized Losses | *RL* | The number of losing stocks among the stocks sold on trading days with sell transactions |
| Paper Gains | *PG* | The number of profitable stocks among the stocks still held on trading days with sell transactions |
| Paper Losses | *PL* | The number of losing stocks among the stocks still held on trading days with sell transactions |
| Social Media Information | *Num* | Logarithm of one plus the post volume by users followed by investors |
| Postive Information | *Pos* | Logarithm of one plus the positive post volume by users followed by investors |
| Negative Information | *Neg* | Logarithm of one plus the negative post volume by users followed by investors |
| Neutral Information | *Neu* | Logarithm of one plus the neutral post volume by users followed by investors |
| Market Return | *MktRet* | Logarithmic returns of the CSI 300 Index |
| Trading Frequency | *TransNum* | Logarithm of one plus the number of trades by investors |
| Number of Stocks Held | *StkNum* | Logarithm of one plus the number of stocks held by investors |
| Holding Price | *StkPrice* | Logarithm of one plus the average price of stocks held by investors |

# 3.3. Model Specification

The main objective of this study is to examine the impact of information obtained by individual investors from social media on the disposition effect. To achieve this, we have constructed the following regression model:

$$DE_{i,t} = \beta_0 + \beta_1 Num_{i,[t-3,t]} + \beta_2 Controls_{i,[t-3,t]} + Individual_i + \varepsilon_{i,t}$$

In this model, the dependent variable $DE_{i,t}$ is the disposition effect of individual investor *i* on trading day *t*; the independent variable $Num_{i,[t-3,t]}$ is the information obtained by investors from social media in the three days prior to trading. $Controls_{i,[t-3,t]}$ represent the market return, trading frequency, number of stocks held, and average holding price in the three days prior to trading, respectively. $Individual_i$ denotes the fixed effects for individual investors, and $\varepsilon_{i,t}$ represents the regression residuals. This study focuses on the regression coefficient $\beta_1$ of the independent variable $Num_{i,[t-3,t]}$ : if $\beta_1$ is significantly greater than 0, it indicates that social media information increases the disposition effect of individual investors; if $\beta_1$ is significantly less than 0, it indicates that social media information decreases the disposition effect of individual investors.

# 4. Empirical Results

## 4.1. Descriptive Statistics

To observe the overall trading status of the Xueqiu real trading portfolio users, this paper uses the portfolio net value data provided by Xueqiu, which covers nearly 800 trading days for each real portfolio. The daily logarithmic average return of the portfolios within the sample is calculated and compared with the daily logarithmic return of the CSI 300 index, as shown in Figure 1. From Figure 1, it can be observed that, except for a few days when the returns of the Xueqiu real portfolios are higher or lower, the returns of the Xueqiu real portfolios mostly fluctuate within the range of the CSI 300 index returns. Therefore, it can be concluded that the data of the Xueqiu real portfolios within the sample do not show significant deviation and are usable.

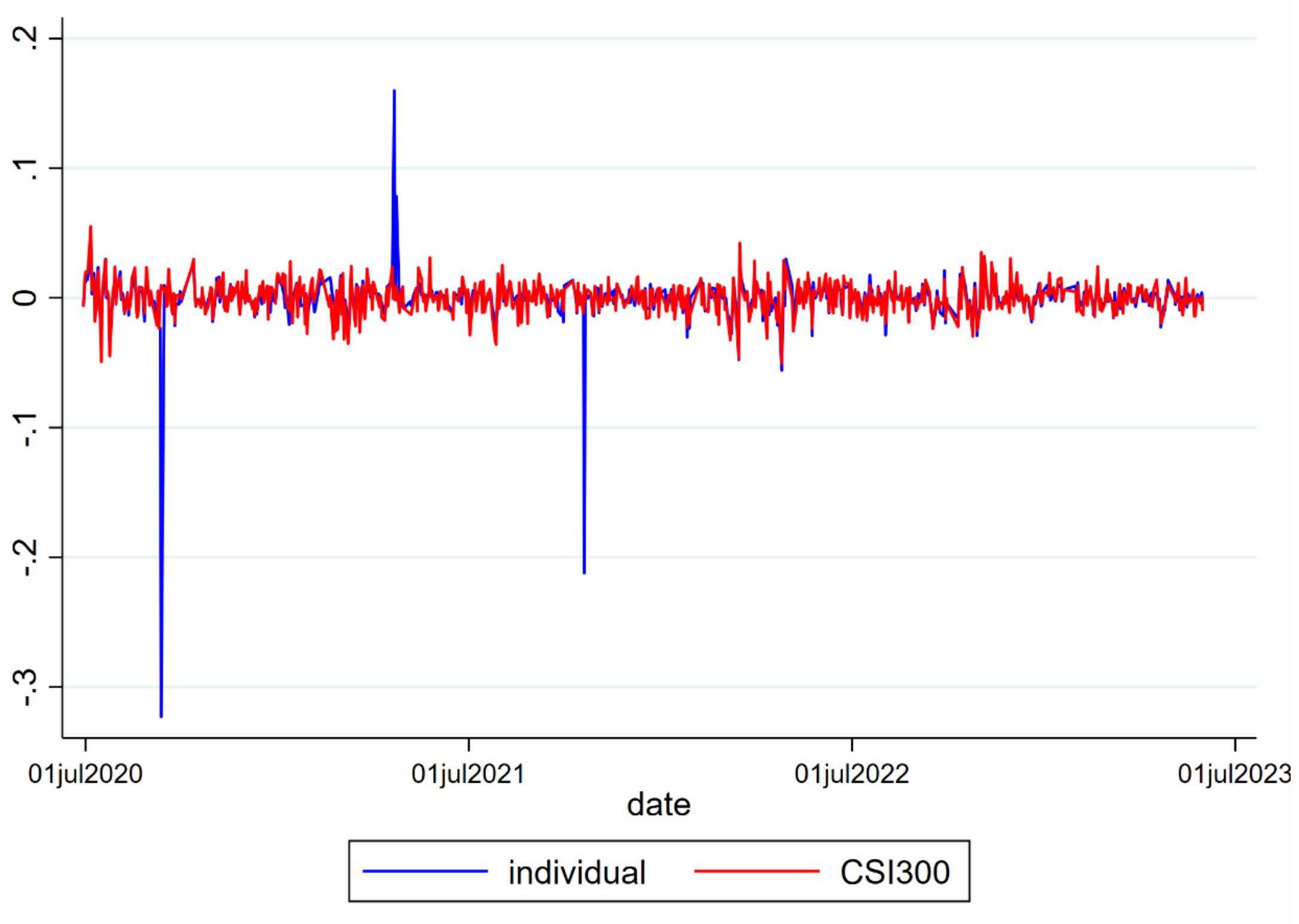

**Figure 1.** Time Series Plot of Xueqiu Real Portfolio Returns and CSI 300 Index Returns

Table 2 presents the descriptive statistics for each variable. During the sample period, the average disposition effect for individual investors is 0.1203, indicating a tendency to sell profitable stocks too early while holding onto losing stocks, which reflects the presence of the disposition effect. The average proportion of realized gains is 0.4756, and the average proportion of realized losses is 0.3553. Compared to the disposition effect, the values for social media information, positive information, negative information, neutral information, trading frequency, number of stocks held, and holding price are relatively higher. Consequently, logarithmic transformations were applied to these variables in the study. Table 2 presents the statistical results following these transformations.

**Table 2.** Summary statistics for the variables (129,865 Observations).

| Variable | Mean | St.dev | Min | 25th Pct | 50th Pct | 75th Pct | Max |
|---|---|---|---|---|---|---|---|
| *DE* | 0.1203 | 0.7688 | -1 | -0.5 | 0.2857 | 1 | 1 |
| *PGR* | 0.4756 | 0.4456 | 0 | 0 | 0.5 | 1 | 1 |
| *PLR* | 0.3553 | 0.4215 | 0 | 0 | 0 | 1 | 1 |
| *RG* | 0.7725 | 0.8466 | 0 | 0 | 1 | 1 | 22 |
| *RL* | 0.6424 | 0.8596 | 0 | 0 | 0 | 1 | 40 |
| *PG* | 0.5476 | 1.181 | 0 | 0 | 0 | 1 | 30 |
| *PL* | 0.9517 | 1.5872 | 0 | 0 | 0 | 1 | 36 |
| *Num* | 4.742 | 2.0226 | 0 | 3.3673 | 4.7449 | 6.4394 | 10.0071 |
| *Pos* | 4.2145 | 1.9622 | 0 | 2.8904 | 4.1431 | 5.8972 | 9.3236 |
| *Neg* | 3.3656 | 1.9674 | 0 | 1.7918 | 3.3322 | 5.0039 | 8.8989 |
| *Neu* | 2.3907 | 1.6514 | 0 | 1.0986 | 2.3026 | 3.7377 | 7.5143 |
| *MktRet* | 0.0011 | 0.0165 | -0.0821 | -0.007 | 0.0014 | 0.01 | 0.0771 |
| *TransNum* | 0.8045 | 0.8284 | 0 | 0 | 0.6931 | 1.3863 | 4.1431 |
| *StkNum* | 1.1979 | 0.4993 | 0.2877 | 0.6931 | 1.0986 | 1.5404 | 3.8286 |
| *StkPrice* | 2.9081 | 0.7528 | 0.1655 | 2.4014 | 2.8748 | 3.3708 | 7.6402 |

## 4.2. Benchmark Regression

### 4.2.1. Impact of social media information on disposition effect

Table 3 presents the regression results on the impact of information from social media on the disposition effect among individual investors. The results indicate that, after controlling for market returns, trading frequency, number of stocks held, and holding prices, and accounting for individual fixed effects, the coefficient for Num is -0.0128 and is significant at the 1% level. This suggests that social media information significantly reduces the disposition effect among individual investors, supporting research hypothesis 1.

**Table 3.** Regressions of disposition effect on social media information.

| Variable | *DE* |
|---|---|
| *Num* | -0.0128***<br>(-3.44) |
| *MktRet* | 4.3530***<br>(31.95) |
| *TransNum* | 0.0373***<br>(10.32) |
| *StkNum* | -0.0136* |
| | (-1.74) |
| *StkPrice* | 0.0519***<br>(10.46) |
| *Constant* | 0.0117<br>(0.49) |
| *Individual* | Fixed |
| *Observations* | 129865 |
| *R-squared* | 0.0120 |

Note: The values in parentheses are the t-values of the coefficients, with standard errors clustered at the individual investor level. ***, **, and * denote significance levels of 1%, 5%, and 10%, respectively.

### 4.2.2. Impact of social media information on the various components of the disposition effect

Although the regression results in Table 3 show that social media information can significantly reduce the disposition effect among individual investors, the underlying reasons remain unclear. Therefore, this study will further explore how social media information affects the disposition effect. According to Odean's [3] definition, the disposition effect consists of the difference between the proportion of realized gains and the proportion of realized losses. The proportion of realized gains is calculated from realized gains and paper gains, while the proportion of realized losses is calculated from realized losses and paper losses. To better understand the impact of social media information on each component of the disposition effect, this research will independently regress these factors, with the results presented in Table 4.

**Table 4.** Regressions of *PGR*, *PLR*, *RG*, *PG*, *RL*, and *PL* on social media information.

| Variable | *PGR* | *PLR* | *RG* | *PG* | *RL* | *PL* |
|---|---|---|---|---|---|---|
| *Num* | -0.0128***<br>(-5.91) | $-5.41\times10^{-6}$<br>(-0.00) | -0.0234***<br>(-6.04) | 0.0155***<br>(2.91) | 0.0059<br>(1.5) | 0.0246***<br>(3.88) |
| *MktRet* | 2.2761***<br>(28.9) | -2.0769***<br>(-29.08) | 5.2610***<br>(35.13) | 5.2488***<br>(27.08) | -5.4163***<br>(-33.09) | -5.9729***<br>(-26.99) |
| *TransNum* | 0.0506*** | 0.0133*** | 0.1027*** | - | 0.0116** | -0.1852*** |
| | (24.22) | (6.76) | (22.98) | 0.0433***<br>(-9.6) | (2.54) | (-29.46) |
| *StkNum* | -0.0752***<br>(-15.53) | -0.0616***<br>(-13.84) | 0.2606***<br>(24.54) | 0.9187***<br>(33.24) | 0.3281***<br>(27.48) | 1.8358***<br>(50.36) |
| *StkPrice* | 0.0295***<br>(10.41) | -0.0224***<br>(-8.71) | 0.0434***<br>(8.96) | 0.0065<br>(0.9) | -0.0684***<br>(-14.17) | -0.1597***<br>(-19.07) |
| *Constant* | 0.4975***<br>(35.8) | 0.4858***<br>(38.79) | 0.3565***<br>(14.68) | -0.6161***<br>(-15.41) | 0.4170***<br>(15.97) | -0.7441***<br>(-15.98) |
| *Individual* | Fixed | Fixed | Fixed | Fixed | Fixed | Fixed |
| *Observations* | 129865 | 129865 | 129865 | 129865 | 129865 | 129865 |
| *R-squared* | 0.0169 | 0.0118 | 0.0341 | 0.1281 | 0.0292 | 0.2761 |

Note: The values in parentheses are the t-values of the coefficients, with standard errors clustered at the individual investor level. ***, **, and * denote significance levels of 1%, 5%, and 10%, respectively.

The results in Table 4 reveal that the regression coefficient of *Num* on *PGR* is significantly negative, while the coefficient of *Num* on *PLR* is not significant. This suggests that as individual investors receive more information from social media, the proportion of realized gains significantly decreases, whereas the proportion of realized losses remains relatively stable. Additionally, the regression coefficient of *Num* on *RG* is significantly negative, and the coefficient of *Num* on *PG* is significantly positive. These findings further confirm that as investors receive more information from social media, they are more likely to retain profitable stocks rather than selling them immediately. Although *Num* does not significantly affect *RL*, the coefficient of *Num* on *PL* is significantly positive, indicating that investors are inclined to hold onto losing stocks rather than selling them as they acquire more information from social media.

Additionally, from the coefficients and significance levels of the control variables in Tables 3 and 4, it can be observed that as the market return increases, individual investors tend to sell profitable stocks and hold losing stocks, which exacerbates the disposition effect. The more frequently individual investors trade, the more irrational their trading behavior becomes, leading to a stronger disposition effect. When the price of the stocks held is higher, trading behavior becomes more irrational, the phenomenon of selling winners and holding losers becomes more pronounced, and the disposition effect becomes stronger. As the number of stocks held increases, trading behavior becomes more rational, and the disposition effect decreases, though the effect is not very significant.

## 4.3. Robustness Tests

### 4.3.1. Recalculate the disposition effect based on selling all positions

In the previous section, this paper calculated the disposition effect for each selling day, including both partial sales and the sale of all stocks. Statistics show that 66.27% of all sell transactions represented selling all of the stock, and 79.50% of the first sell transactions were full sales. In the robustness test, this paper follows the approach of Odean [3] and Wu et al. [2] by recalculating the disposition effect based solely on selling all positions. The regression results, presented in column (1) of Table 5, indicate that, when focusing solely on selling all positions and excluding partial sales, the coefficient of *Num* remains significantly negative at the 1% level. This suggests that the finding that social media information significantly reduces investors' disposition effect is robust.

### 4.3.2. Recalculate the disposition effect based on percentage differences

In the previous analysis of the disposition effect, the absolute difference between *PGR* and *PLR*

(*PGR*-*PLR*) was used. Following the approach of Wu et al. [2], the disposition effect is now measured using the percentage difference (*PGR*/*PLR* - 1). The regression results, presented in column (2) of Table 5, show that after recalculating the disposition effect based on the percentage difference, the coefficient of *Num* remains significantly negative at the 1% level. This suggests that the finding that social media information significantly reduces investors' disposition effect is robust.

### 4.3.3. Consider the post-holiday effect

Trading is not possible during market closures for holidays, so any information accumulated during this period may influence stock trading on the first trading day after the holiday. This study posits that the disposition effect of investors on this first trading day may differ from that on other trading days. To account for the post-holiday effect, a dummy variable, *PostHoliday*, is included in the regression. Weekends, like holidays, also exhibit this effect, so they are treated as holidays in this study. The dummy variable *PostHoliday* is set to 1 for the first trading day after the holiday and 0 otherwise. The regression results, presented in column (3) of Table 5, show that after controlling for the post-holiday effect, the coefficient of *Num* is significantly negative at the 5% level, confirming that social media information significantly reduces the disposition effect. Additionally, the results reveal that the disposition effect of individual investors on the first trading day after the holiday is 0.0230 higher compared with other trading days, with this difference being significant at the 1% level.

**Table 5.** Robustness test.

| **Variable** | **(1)**<br>***DE*** | **(2)**<br>***DE*** | **(3)**<br>***DE*** |
|---|---|---|---|
| *Num* | -0.0131***<br>(-2.96) | -0.0155***<br>(-3.9) | -0.0093**<br>(-2.44) |
| *MktRet* | 4.6199***<br>(27.4) | 1.2143***<br>(7.57) | 4.3707***<br>(32.07) |

| *TransNum* | 0.0316***<br>(7.29) | 0.0664***<br>(15.46) | 0.0401***<br>(10.71) |
|---|---|---|---|
| *StkNum* | -0.0336***<br>(-3.73) | 0.4352***<br>(35.84) | -0.0156**<br>(-1.98) |
| *StkPrice* | 0.0582***<br>(10.4) | -0.0136***<br>(-3.3) | 0.0515***<br>(10.37) |
| *AfterHoliday* | | | 0.0230***<br>(4.29) |
| *Constant* | 0.0289<br>(1.06) | -1.2118***<br>(-50.83) | -0.0088<br>(-0.36) |
| *Individual* | Fixed | Fixed | Fixed |
| *Observations* | 95673 | 63428 | 129865 |
| *R-squared* | 0.0116 | 0.0555 | 0.0122 |

Note: The values in parentheses are the t-values of the coefficients, with standard errors clustered at the individual investor level. ***, **, and * denote significance levels of 1%, 5%, and 10%, respectively.

## 4.4. Mechanism analysis

Considering that social media posts reflect the poster's bullish or bearish outlook on stocks and that retail investors' sentiment can influence both stock returns [32] and trading volume [33], this paper posits that the sentiment of posts investors focus on can affect their expectations of future stock returns, thereby influencing the disposition effect. Thus, the paper aims to explore this impact mechanism from the perspective of post sentiment.

For text sentiment classification, common methods include machine learning and sentiment dictionaries. Machine learning methods require training data with subjective sentiment judgments, which can directly affect classification results. In contrast, sentiment dictionary methods are more objective. Therefore, this paper uses sentiment dictionaries to classify post sentiments. Given that most posts are made by individual investors and vary widely in form and content, a sentiment dictionary based on a corpus of social media posts is needed. Additionally, since users on social media also share financial news, a sentiment dictionary based on financial news corpus is also required. This paper follows the approach of Fan et al. [34] by combining their sentiment dictionary, which is based on stock forum posts, with the Chinese financial sentiment dictionary developed by Jiang et al. [35], which is based on Chinese financial news texts, for classification.

The specific steps for sentiment classification of post texts in this study are as follows: First, collect posts from users followed by individual investors over the three days prior to their trades, resulting in 9,944,214 posts. Second, clean the text by removing redundant spaces, fields, invalid symbols, emojis, and images. Third, use Python's jieba library for word segmentation and remove stop words. After that, use a sentiment dictionary to identify sentiment words in the text. Considering that negation words can alter sentiment polarity, and following the approach of Jiang et al. [35], we assume that a sentiment word is only influenced by the preceding negation word. Thus, the negation word before the sentiment word and the sentiment word itself are treated as a single sentiment unit. Finally, positive words are assigned a weight of 1, and negative words a weight of -1. Considering the case where double negation implies affirmation, the weight of negation words is set to $(-1)^n$, where *n* is the number of occurrences of the negation word. Thus, the formula for calculating the sentiment score of a single post is as follows:

$$Sentiment = \frac{\sum_{i=1}^{N} (-1)^n \times W_i}{N},$$

$$W_i = \begin{cases} 1, positive \\ -1, negative \end{cases}$$

In this, *Sentiment* represents the emotional score of an individual post, $W_i$ is the weight of sentiment words, and *N* is the total number of sentiment units in the post. When *Sentiment* is greater than 0, the text is considered positive; when *Sentiment* is less than 0, the text is considered negative;

and when *Sentiment* equals 0, the text is considered neutral.

After classifying the sentiment of posts using the above method, we obtained the number of positive, negative, and neutral posts from the users that individual investors follow in the three days prior to their trades. These numbers were log-transformed and represented by the variables *Pos*, *Neg*, and *Neu*. These were then regressed on *DE*, *PGR*, and *PLR*, with results shown in Table 6. Since neutral posts do not express a clear emotional tendency, Table 6 provides the regression results excluding neutral posts for comparison.

**Table 6.** Regressions of *DE*, *PGR*, and *PLR* on sentiment of social media information.

| **Variable** | ***DE*** | ***DE*** | ***PGR*** | ***PGR*** | ***PLR*** | ***PLR*** |
| --- | --- | --- | --- | --- | --- | --- |
| *Pos* | 0.0072<br>(1.47) | 0.0067<br>(1.3) | -0.0004<br>(-0.14) | -0.0005<br>(-0.18) | -0.0076***<br>(-2.92) | -0.0072***<br>(-2.64) |
| *Neg* | -0.0274***<br>(-5.14) | -0.0283***<br>(-4.98) | -0.0179***<br>(-5.86) | -0.0181***<br>(-5.55) | 0.0096***<br>(3.32) | 0.0102***<br>(3.33) |
| *Neu* | | 0.0022<br>(0.39) | | 0.0006<br>(0.18) | | -0.0016<br>(-0.54) |
| *MktRet* | 4.2526***<br>(30.8) | 4.2522***<br>(30.8) | 2.2193***<br>(27.81) | 2.2192***<br>(27.81) | -2.0333***<br>(-28.12) | -2.0330***<br>(-28.12) |
| *TransNum* | 0.0376***<br>(10.36) | 0.0375***<br>(10.34) | 0.0509***<br>(24.29) | 0.0509***<br>(24.25) | 0.0133***<br>(6.76) | 0.0134***<br>(6.77) |
| *StkNum* | -0.0138*<br>(-1.77) | -0.0138*<br>(-1.77) | -0.0754***<br>(-15.57) | -0.0754***<br>(-15.57) | -0.0615***<br>(-13.82) | -0.0615***<br>(-13.82) |
| *StkPrice* | 0.0517***<br>(10.42) | 0.0517***<br>(10.42) | 0.0294***<br>(10.37) | 0.0294***<br>(10.37) | -0.0223***<br>(-8.68) | -0.0223***<br>(-8.68) |
| *Constant* | 0.0133<br>(0.58) | 0.0135<br>(0.59) | 0.4988***<br>(37.12) | 0.4988***<br>(37.16) | 0.4855***<br>(40.25) | 0.4853***<br>(40.25) |
| *Individual* | Fixed | Fixed | Fixed | Fixed | Fixed | Fixed |
| *Observations* | 129865 | 129865 | 129865 | 129865 | 129865 | 129865 |
| *R-squared* | 0.0122 | 0.0122 | 0.0171 | 0.0171 | 0.0119 | 0.0119 |

Note: The values in parentheses are the t-values of the coefficients, with standard errors clustered at the individual investor level. ***, **, and * denote significance levels of 1%, 5%, and 10%, respectively.

Table 6 shows that information with different emotional tendencies on social media affects the disposition effect in varying ways. Negative information can significantly reduce the disposition effect, while positive and neutral information have no significant impact on the disposition effect. Specifically, positive information has a significantly negative effect on *PLR*, indicating that when investors focus on positive information, they tend to hold onto losing stocks in hopes of recouping losses, thereby exacerbating the disposition effect. Conversely, negative information has a significantly negative effect on *PGR* and a significantly positive effect on *PLR*, suggesting that when investors focus on negative information, they are more likely to hold onto winning stocks while selling losing ones to cut losses, thereby reducing the disposition effect. Overall, positive information has no significant impact on the disposition effect, while negative information significantly affects the disposition effect. In other words, it is through negative social media information that the disposition effect is reduced, leading investors to make more rational adjustments to their portfolios.

The impact of different emotional information on the disposition effect on social media is not the same. When investors receive positive information, they tend to have overly optimistic expectations about future stock returns, leading them to hold onto losing stocks in hopes of recovering their losses, while their decision to sell winning stocks is not significantly affected. When negative information is received, investors become more rational and cautious in their expectations for future returns. They worry that losing stocks may continue to decline, so they tend to sell them early, while still holding onto winning stocks for further observation. Overall, negative information has a large and significant impact on the disposition effect, thus social media information reduces the disposition effect.

## 4.4. Heterogeneity Analysis

This study has demonstrated that social media information can reduce the disposition effect through negative information. Previous research has shown that individual investor characteristics, such as investment experience [9], region [5], and gender [5, 9, 12], can affect the disposition effect. However, it remains unclear whether the impact of negative information in social media on the disposition effect varies with different investor characteristics. Using data from actual trading accounts, this study groups investors by investment experience, region, and gender to examine heterogeneity. Additionally, since social media information affects the disposition effect and different investors follow different users, the impact of information from various users may vary. Therefore, in addition to the heterogeneity tests mentioned above, this study also conducts a grouped regression analysis based on the differences in the users followed by individual investors.

### 4.4.1. Differences in investment experience

The findings indicate that individual investors benefit from negative information acquired through social media, which helps reduce their disposition effect. However, the ability to form cognitive insights and expectations based on information from social media may vary among individual investors with different levels of experience. Previous literature has used the number of years an investor has traded [36-38], the total number of stocks held [38], and the total number of trades [38] to gauge investment experience. Based on the data obtained in this study, we categorized investors according to the median number of followers, the number of stocks followed, and the number of posts made. Investors with values above the median were classified as experienced investors, while those below the median were classified as inexperienced investors. Results from Table 7 show that when using the number of followers, the number of stocks followed, and the number of posts as indicators of investment experience, the coefficient of Num is significantly strong in the group with rich investment experience. Furthermore, there is significant heterogeneity between the group with rich investment experience and the group with insufficient experience, indicating notable differences in the ability to form cognitive insights and expectations based on social media information between investors with high and low levels of experience. Furthermore, the coefficient of Neg shows strong significance in all groups. In the group with more investment experience, the absolute value of the Neg coefficient is larger, indicating that in the group with more investment experience, the impact of negative information on the disposition effect is stronger. For investors with less investment experience, the impact of negative information on the disposition effect is relatively small, which results in the overall insignificance of social media information on the disposition effect. Overall, there is no significant difference in the Neg coefficients between the two groups.

**Table 7.** Heterogeneity test among individual investors with different investment experiences.

**Panel A** Measures investment experience by the number of followers.

| **Variable** | **group with high investment experience** | **group with low investment experience** | **group with high investment experience** | **group with low investment experience** | **group with high investment experience** | **group with low investment experience** |
|---|---|---|---|---|---|---|
| | ***DE*** | ***DE*** | ***DE*** | ***DE*** | ***DE*** | ***DE*** |
| *Num* | -0.0201*** (-3.61) | -0.0073 (-1.47) | | | | |
| *Pos* | | | 0.007 (0.89) | 0.0082 (1.29) | 0.0057 (0.69) | 0.0076 (1.16) |
| *Neg* | | | -0.0360*** (-4.09) | -0.0210*** (-3.14) | -0.0385*** (-4.02) | -0.0219*** (-3.1) |

| *Neu* | | | | | 0.0052<br>(0.59) | 0.0023<br>(0.31) |
|---|---|---|---|---|---|---|
| *MktRet* | 4.3868***<br>(23.12) | 4.3082***<br>(22) | 4.2577***<br>(21.94) | 4.2285***<br>(21.42) | 4.2559***<br>(21.97) | 4.2282***<br>(21.42) |
| *TransNum* | 0.0332***<br>(6.65) | 0.0419***<br>(8.02) | 0.0338***<br>(6.73) | 0.0419***<br>(8) | 0.0337***<br>(6.7) | 0.0419***<br>(7.99) |
| *StkNum* | -0.0166<br>(-1.54) | -0.0105<br>(-0.93) | -0.0171<br>(-1.57) | -0.0106<br>(-0.93) | -0.017<br>(-1.57) | -0.0106<br>(-0.94) |
| *StkPrice* | 0.0520***<br>(7.43) | 0.0518***<br>(7.36) | 0.0520***<br>(7.41) | 0.0515***<br>(7.33) | 0.0519***<br>(7.41) | 0.0515***<br>(7.33) |
| *Constant* | 0.0575<br>(1.54) | -0.0166<br>(-0.55) | 0.0616*<br>(1.71) | -0.0193<br>(-0.66) | 0.0628*<br>(1.74) | -0.0191<br>(-0.65) |
| *Individual* | Fixed | Fixed | Fixed | Fixed | Fixed | Fixed |
| *Observations* | 66755 | 63110 | 66755 | 63110 | 66755 | 63110 |
| *R-squared* | 0.0123 | 0.0119 | 0.0125 | 0.0120 | 0.0125 | 0.0121 |
| coefficient differences between groups of Num | 0.013*** | | | | | |
| coefficient differences between groups of Pos | | | 0.001 | | 0.002 | |
| coefficient differences between groups of Neg | | | 0.015 | | 0.017** | |
| coefficient differences between groups of Neu | | | | | -0.003* | |

**Panel B** Measures investment experience by the number of stocks followed

| **Variable** | **group with high investment experience** | **group with low investment experience** | **group with high investment experience** | **group with low investment experience** | **group with high investment experience** | **group with low investment experience** |
|---|---|---|---|---|---|---|
| | ***DE*** | ***DE*** | ***DE*** | ***DE*** | ***DE*** | ***DE*** |
| *Num* | -0.0157***<br>(-2.94) | -0.0097*<br>(-1.87) | | | | |
| *Pos* | | | 0.0105<br>(1.48) | 0.0052<br>(0.76) | 0.0103<br>(1.41) | 0.004<br>(0.55) |
| *Neg* | | | -0.0352***<br>(-4.46) | -0.0204***<br>(-2.84) | -0.0355***<br>(-4.07) | -0.0222***<br>(-2.96) |
| *Neu* | | | | | 0.0006<br>(0.08) | 0.0049<br>(0.62) |
| *MktRet* | 4.1449***<br>(22.64) | 4.5969***<br>(22.52) | 4.0131***<br>(21.66) | 4.5232***<br>(21.86) | 4.0128***<br>(21.66) | 4.5230***<br>(21.87) |
| *TransNum* | 0.0363***<br>(7.61) | 0.0382***<br>(6.96) | 0.0367***<br>(7.65) | 0.0384***<br>(6.97) | 0.0367***<br>(7.64) | 0.0383***<br>(6.93) |
| *StkNum* | -0.0193*<br>(-1.91) | -0.0062<br>(-0.5) | -0.0194*<br>(-1.93) | -0.0064<br>(-0.52) | -0.0194*<br>(-1.93) | -0.0065<br>(-0.52) |
| *StkPrice* | 0.0465***<br>(7.05) | 0.0574***<br>(7.72) | 0.0464***<br>(7.02) | 0.0572***<br>(7.69) | 0.0464***<br>(7.02) | 0.0572***<br>(7.69) |
| *Constant* | 0.0696*<br>(1.95) | -0.0478<br>(-1.51) | 0.0744**<br>(2.18) | -0.0497<br>(-1.62) | 0.0746**<br>(2.18) | -0.0496<br>(-1.62) |
| *Individual* | Fixed | Fixed | Fixed | Fixed | Fixed | Fixed |
| *Observations* | 69509 | 60356 | 69509 | 60356 | 69509 | 60356 |
| *R-squared* | 0.0115 | 0.0127 | 0.0118 | 0.0128 | 0.0118 | 0.0128 |
| coefficient differences between groups of Num | 0.006* | | | | | |
| coefficient differences between groups of Pos | | | -0.005 | | -0.006 | |
| coefficient differences between groups of Neg | | | 0.015 | | 0.013* | |
| coefficient differences between groups of Neu | | | | | 0.004 | |

**Panel C** Measures investment experience by the number of posts.

| **Variable** | **group with high investment experience** | **group with low investment experience** | **group with high investment experience** | **group with low investment experience** | **group with high investment experience** | **group with low investment experience** |
|---|---|---|---|---|---|---|
| | ***DE*** | ***DE*** | ***DE*** | ***DE*** | ***DE*** | ***DE*** |
| *Num* | - | -0.0051 | | | | |
| | 0.0210***<br>(-3.83) | (-1.02) | | | | |
| *Pos* | | | 0.0028<br>(0.37) | 0.0113*<br>(1.76) | 0.0032<br>(0.4) | 0.0096<br>(1.43) |
| *Neg* | | | -0.0317***<br>(-3.81) | -0.0223***<br>(-3.21) | -0.0310***<br>(-3.41) | -0.0249***<br>(-3.42) |
| *Neu* | | | | | -0.0014<br>(-0.17) | 0.007<br>(0.9) |
| *MktRet* | 4.1787***<br>(22.81) | 4.5402***<br>(22.38) | 4.0735***<br>(21.74) | 4.4491***<br>(21.79) | 4.0739***<br>(21.75) | 4.4486***<br>(21.78) |
| *TransNum* | 0.0431***<br>(8.75) | 0.0306***<br>(5.77) | 0.0436***<br>(8.81) | 0.0306***<br>(5.76) | 0.0436***<br>(8.81) | 0.0304***<br>(5.72) |
| *StkNum* | -0.0261**<br>(-2.48) | 0.0013<br>(0.11) | -0.0263**<br>(-2.5) | 0.0011<br>(0.09) | -0.0263**<br>(-2.5) | 0.0011<br>(0.09) |
| *StkPrice* | 0.0528***<br>(7.89) | 0.0509***<br>(6.89) | 0.0525***<br>(7.85) | 0.0507***<br>(6.87) | 0.0525***<br>(7.85) | 0.0507***<br>(6.87) |
| *Constant* | 0.0769**<br>(2.1) | -0.0448<br>(-1.45) | 0.0772**<br>(2.19) | -0.0452<br>(-1.52) | 0.0769**<br>(2.19) | -0.0451<br>(-1.51) |
| *Individual* | Fixed | Fixed | Fixed | Fixed | Fixed | Fixed |
| *Observations* | 70370 | 59495 | 70370 | 59495 | 70370 | 59495 |
| *R-squared* | 0.0117 | 0.0126 | 0.0119 | 0.0128 | 0.0119 | 0.0128 |
| coefficient differences between groups of Num | 0.016*** | | | | | |
| coefficient differences between groups of Pos | | | 0.008* | | 0.006* | |
| coefficient differences between groups of Neg | | | 0.009 | | 0.006 | |
| coefficient differences between groups of Neu | | | | | 0.008 | |

Note: The values in parentheses are the t-values of the coefficients, with standard errors clustered at the individual investor level. ***, **, and * denote significance levels of 1%, 5%, and 10%, respectively.

### 4.4.2. Differences in users followed

Opinion leaders are influential figures who shape opinions during information dissemination or interpersonal interactions. Previous studies have shown that the opinions of opinion leaders in stock forums can widely circulate among investors and impact their behavior [39]. Therefore, this paper hypothesizes that there are likely differences in the impact of information from opinion leaders versus non-opinion leaders on investors' disposition effect. In this study, users who are followed by individual investors within the sample are ranked by their follower count, with the top 1% being classified as opinion leaders. Investors are divided into two groups based on the median proportion of opinion leaders they follow: those who follow more opinion leaders and those who follow fewer. The results of the regression analysis are presented in Panel A of Table 8.

For investors who follow more opinion leaders and those who follow fewer opinion leaders, the coefficients of Num are both highly significant, indicating that after receiving social media information, the disposition effect is significantly reduced. Furthermore, the coefficients of Neg are highly significant for both groups of investors, indicating that each group can reduce its disposition effect through negative information. Compared to investors who follow more opinion leaders, those who follow fewer can reduce the disposition effect more effectively through social media information, particularly by using negative information. Existing research has shown that investors tend to focus on opinion leaders who express positive views while overlooking the accuracy of these views, and often dismiss opinion leaders who can accurately predict market downturns[39]. Therefore, the possible explanation for the conclusion that investors who follow fewer opinion leaders can better utilize social media information to reduce the disposition effect is that investors who follow more opinion leaders are prone to rely on the positive information shared by these leaders, neglecting the negative aspects and lacking rational analysis of the information. In contrast, investors who follow fewer opinion leaders are less likely to blindly follow positive information and more likely to consider negative information, enabling them to make more rational judgments and decisions. Consequently, compared to those who follow more opinion leaders, investors who follow fewer are better able to leverage social media information to reduce the disposition effect to a greater extent and are more strongly influenced by negative information. Overall, the difference between the two groups is not significant.

### 4.4.3. Differences in regions

Compared to investors in non-first-tier cities, those in first-tier cities (Beijing, Shanghai, Guangzhou, Shenzhen) may have more financial knowledge and greater information advantages. Consequently, investors from different regions might react differently to information obtained from social media. To explore this heterogeneity, this paper conducts a grouped regression based on whether individual investors are from first-tier cities or not, with results shown in Panel B of Table 8. For investors in first-tier cities, social media information significantly reduces the disposition effect. Further research shows that the impact of negative information on the disposition effect is negative but not significant. For investors in non-first-tier cities, the impact of negative information on their disposition effect is significantly negative, but overall, the effect of social media information on the disposition effect is not significant. This may be due to the inconsistent effect of positive information on the disposition effect for different investor groups, which leads to differences in overall significance. In general, there is no significant difference between the two groups.

### 4.4.4. Differences in gender

Compared to male investors, female investors generally have lower levels of financial knowledge[10], higher risk aversion[11], and are more influenced by regret[12]. As a result, female and male investors may make different decisions after receiving information from social media. This study examines heterogeneity based on the gender of individual investors through grouped regression, with results shown in Panel C of Table 8. For male investors, social media information significantly reduces the disposition effect, whereas for female investors, social media information does not have a significant impact on the disposition effect. Furthermore, for male investors, negative information on social media significantly reduces their disposition effect, while the effect of positive information is not significant. For female investors, neither positive nor negative information has a significant effect on the disposition effect. Compared to male investors, female investors might not extract useful information from social media due to their generally lower financial knowledge[10], which could result in no significant reduction in the disposition effect. Additionally, social media information might not significantly alter their risk aversion[11] or susceptibility to regret[12], leading them to continue selling winners and holding losers.

According to the results of the inter-group difference test, there is no significant difference in the impact of social media information on male and female investors. However, the impact of negative information on these two types of investors shows a significant difference. This may be because negative information on social media reduces the disposition effect for male investors, while positive information increases their disposition effect. In contrast, female investors' disposition effect is not significantly influenced by social media information. As a result, overall, there is no significant difference in the impact of social media information on the disposition effect between genders.

**Table 8.** Heterogeneity test among individual investors with different characteristics.

**Panel A** Heterogeneity of individual investors based on the differences in users followed.

| Variable | Investors who follow a large number of opinion leaders | Investors who follow a small number of opinion leaders | Investors who follow a large number of opinion leaders | Investors who follow a small number of opinion leaders | Investors who follow a large number of opinion leaders | Investors who follow a small number of opinion leaders |
|---|---|---|---|---|---|---|
| | ***DE*** | ***DE*** | ***DE*** | ***DE*** | ***DE*** | ***DE*** |
| *Num* | -0.0104**<br>(-2.02) | -0.0148***<br>(-2.75) | | | | |
| *Pos* | | | 0.0093<br>(1.36) | 0.0058<br>(0.82) | 0.0099<br>(1.37) | 0.0043<br>(0.57) |
| *Neg* | | | -0.0262***<br>(-3.65) | -0.0287***<br>(-3.63) | -0.0254***<br>(-3.32) | -0.0319***<br>(-3.77) |
| *Neu* | | | | | -0.0022<br>(-0.28) | 0.0069<br>(0.84) |
| *MktRet* | 4.4017***<br>(21.89) | 4.3080***<br>(23.26) | 4.3037***<br>(21.18) | 4.2042***<br>(22.32) | 4.3038***<br>(21.19) | 4.2022***<br>(22.32) |
| *TransNum* | 0.0303***<br>(5.51) | 0.0429***<br>(8.96) | 0.0302***<br>(5.48) | 0.0434***<br>(9.03) | 0.0302***<br>(5.49) | 0.0433***<br>(8.98) |
| *StkNum* | -0.0121<br>(-1.08) | -0.0148<br>(-1.37) | -0.0125<br>(-1.11) | -0.0149<br>(-1.38) | -0.0125<br>(-1.11) | -0.0149<br>(-1.37) |
| *StkPrice* | 0.0543***<br>(7.38) | 0.0498***<br>(7.43) | 0.0542***<br>(7.36) | 0.0496***<br>(7.39) | 0.0542***<br>(7.36) | 0.0496***<br>(7.39) |
| *Constant* | -0.0047<br>(-0.15) | 0.0264<br>(0.76) | -0.0079<br>(-0.26) | 0.0326<br>(0.96) | -0.0081<br>(-0.26) | 0.0334<br>(0.99) |
| *Individual* | Fixed | Fixed | Fixed | Fixed | Fixed | Fixed |
| *Observations* | 59620 | 70245 | 59620 | 70245 | 59620 | 70245 |
| *R-squared* | 0.0120 | 0.0122 | 0.0121 | 0.0123 | 0.0121 | 0.0124 |
| coefficient differences between groups of Num | -0.004 | | | | | |
| coefficient differences between groups of Pos | | | -0.003 | | -0.006 | |
| coefficient differences | | | -0.002 | | -0.006 | |

| between groups of Neg | | | |
|---|---|---|---|
| coefficient differences between groups of Neu | | | 0.009 |

**Panel B** Heterogeneity of individual investors based on the differences in regions.

| **Variable** | **Investors in first-tier cities** | **Investors in non-first-tier cities** | **Investors in first-tier cities** | **Investors in non-first-tier cities** | **Investors in first-tier cities** | **Investors in non-first-tier cities** |
|---|---|---|---|---|---|---|
| | ***DE*** | ***DE*** | ***DE*** | ***DE*** | ***DE*** | ***DE*** |
| *Num* | -0.0262** (-2.12) | -0.011 (-1.17) | | | | |
| *Pos* | | | -0.0116 (-0.72) | 0.0116 (0.84) | -0.0151 (-0.91) | 0.0122 (0.84) |
| *Neg* | | | -0.0211 (-1.14) | -0.0294** (-1.99) | -0.0287 (-1.43) | -0.0283* (-1.77) |
| *Neu* | | | | | 0.0157 (0.82) | -0.0023 (-0.16) |
| *MktRet* | 3.9869*** (10.22) | 4.0563*** (13.55) | 3.9425*** (9.95) | 3.9502*** (12.98) | 3.9364*** (9.93) | 3.9508*** (12.99) |
| *TransNum* | 0.0280*** (2.88) | 0.0335*** (3.72) | 0.0285*** (2.93) | 0.0340*** (3.77) | 0.0281*** (2.89) | 0.0341*** (3.77) |
| *StkNum* | -0.0339 (-1.63) | -0.0112 (-0.6) | -0.0339 (-1.63) | -0.011 (-0.59) | -0.0339 (-1.63) | -0.011 (-0.59) |
| *StkPrice* | 0.0977*** (6.08) | 0.0370*** (3.2) | 0.0977*** (6.07) | 0.0372*** (3.22) | 0.0976*** (6.07) | 0.0372*** (3.22) |
| *Constant* | -0.0327 (-0.41) | 0.0546 (0.85) | -0.034 (-0.43) | 0.057 (0.94) | -0.0315 (-0.4) | 0.0568 (0.94) |
| *Individual* | Fixed | Fixed | Fixed | Fixed | Fixed | Fixed |
| *Observations* | 14297 | 24907 | 14297 | 24907 | 14297 | 24907 |
| *R-squared* | 0.0152 | 0.0101 | 0.0153 | 0.0102 | 0.0153 | 0.0102 |
| coefficient differences between groups of Num | 0.015 | | | | | |
| coefficient differences between groups of Pos | | | 0.023 | | 0.027* | |

| | | | |
|---|---|---|---|
| coefficient differences between groups of Neg | | -0.008 | 0.000 |
| coefficient differences between groups of Neu | | | -0.018 |

**Panel C** Heterogeneity of individual investors based on the differences in gender.

| Variable | Male investors | Female investors | Male investors | Female investors | Male investors | Female investors |
|---|---|---|---|---|---|---|
| | ***DE*** | ***DE*** | ***DE*** | ***DE*** | ***DE*** | ***DE*** |
| *Num* | -0.0169** (-2.02) | -0.017 (-0.74) | | | | |
| *Pos* | | | 0.0082 (0.7) | -0.0289 (-0.8) | 0.0065 (0.53) | -0.023 (-0.56) |
| *Neg* | | | -0.0328*** (-2.59) | 0.006 (0.18) | -0.0361*** (-2.68) | 0.0142 (0.46) |
| *Neu* | | | | | 0.0072 (0.57) | -0.021 (-0.63) |
| *MktRet* | 4.1539*** (15.36) | 4.0704*** (4.2) | 4.0378*** (14.64) | 4.1267*** (4.14) | 4.0354*** (14.64) | 4.1278*** (4.14) |
| *TransNum* | 0.0311*** (4.14) | 0.0440** (2.22) | 0.0315*** (4.18) | 0.0455** (2.28) | 0.0314*** (4.15) | 0.0461** (2.3) |
| *StkNum* | -0.0201 (-1.23) | -0.0792* (-1.85) | -0.0198 (-1.22) | -0.0798* (-1.87) | -0.0199 (-1.22) | -0.0795* (-1.86) |
| *StkPrice* | 0.0587*** (5.64) | 0.0536** (2.17) | 0.0587*** (5.63) | 0.0540** (2.17) | 0.0586*** (5.63) | 0.0533** (2.14) |
| *Constant* | 0.0241 (0.43) | 0.2111 (1.48) | 0.0248 (0.47) | 0.2311* (1.74) | 0.0258 (0.49) | 0.2315* (1.75) |
| *Individual* | Fixed | Fixed | Fixed | Fixed | Fixed | Fixed |
| *Observations* | 31619 | 3107 | 31619 | 3107 | 31619 | 3107 |
| *R-squared* | 0.0116 | 0.0130 | 0.0118 | 0.0133 | 0.0118 | 0.0134 |
| coefficient differences between groups of Num | -0.000 | | | | | |
| coefficient differences between groups of Pos | | | -0.037 | | -0.029 | |
| coefficient differences between groups of | | | 0.039 | | 0.050* | |

| | | | |
|---|---|---|---|
| Neg | | | |
| coefficient differences between groups of Neu | | | -0.028 |

Note: The values in parentheses are the t-values of the coefficients, with standard errors clustered at the individual investor level. ***, **, and * denote significance levels of 1%, 5%, and 10%, respectively.

# 5. Conclusions

This paper uses post data and actual trading data from the mainstream investor social platform Xueqiu to investigate the impact of social media information on individual investors' disposition effect for the first time. The study finds that social media information can significantly reduce the disposition effect. At the individual level, factors such as investment experience, users followed, region, and gender all influence how effectively social media information reduces the disposition effect. Furthermore, this paper delves into the mechanism through which social media information affects the disposition effect. It finds that when receiving positive information, individual investors tend to hold onto losing stocks in hopes of recovering their losses, which exacerbates the disposition effect, though the impact on the disposition effect is not significant. However, when receiving negative information, individual investors are more likely to hold onto winning stocks and sell losing stocks to cut their losses, significantly reducing the disposition effect. In other words, social media information significantly reduces the disposition effect through negative information. Additionally, at the individual level, factors such as investment experience, users followed, region, and gender can all influence the effectiveness of the information acquired by individual investors in reducing the disposition effect.

This study contributes to a deeper understanding of the irrational behavior of individual investors. Individual investors often make investment decisions based on past gains and losses, tending to sell profitable assets too early while holding onto losing assets for too long. Such irrational behavior often impairs the ability to generate returns. Social media information improves the information environment for investors, and exposure to negative information can help make their behavior more rational. Given the high proportion of individual investors in China's stock market, this research is crucial for guiding them towards more scientific and rational investment practices and for promoting the development of the capital market. Of course, this study may have limitations. Since this paper uses trading data from real portfolios on Xueqiu rather than account data provided by brokers, and Xueqiu only publishes the most recent 200 trades for each real portfolio, the sample data may not be comprehensive enough. Moreover, sentiment analysis in Chinese is relatively complex due to the frequent occurrence of sarcasm and irony, which are highly context-dependent. Additionally, a single social media post may convey multiple emotions simultaneously, indicating that the use of sentiment dictionaries for analyzing social media content may have certain limitations. In future research, if we can use a larger sample, expand the sample range to include multiple countries, and apply more precise text analysis methods, the research will be more comprehensive and convincing.

# References

1. Shefrin H, Statman M. The disposition to sell winners too early and ride losers too long: Theory and evidence. The Journal of Finance. 1985;(No.3).
2. Wu J, Wang C, Chen Z, Guo JM. A study of disposition effect among China's individual investors: The perspective of irrational beliefs. Journal of Financial Research. 2020;(02):147-66.
3. Odean T. Are investors reluctant to realize their losses? The Journal of Finance. 1998;(No.5):1775-98.
4. Lu R, Li J, Chen S. Portraits of investors' selling behavior in China's stock market: Advances in disposition effect. Journal of Management World. 2022;38(03):59-78. doi: 10.19744/j.cnki.11-1235/f.2022.0044.
5. Wu Y, Huang W, Su S, Jiang J. Research on the disposition effect of mutual fund investors. Studies of International Finance. 2016;(03):84-96. doi: 10.16475/j.cnki.1006-1029.2016.03.008.
6. Cici G. The prevalence of the disposition effect in mutual funds' trades. Journal of Financial

and Quantitative Analysis. 2012;47(4):795-820. doi: 10.1017/s0022109012000348. PubMed PMID: WOS:000310179000005.
7. Locke PR, Mann SC. Professional trader discipline and trade disposition. Journal of Financial Economics. 2005;76(2):401-44. doi: 10.1016/j.jfineco.2004.01.004. PubMed PMID: WOS:000229066000007.
8. Brown P, Chappel N, Rosa RDS, Walter T. The reach of the disposition effect: Large sample evidence across investor classes. International Review of Finance. 2006;(No.1-2):43-78.
9. Xiao L, Zhao D, Fang Y. An empirical analysis on disposition effect of Chinese margin trading. Chinese Journal of Management Science. 2018;26(09):41-51. doi: 10.16381/j.cnki.issn1003-207x.2018.09.005.
10. Jiang J, Liao L, Wang Z, Xiang H. Financial literacy and retail investors' financial welfare: Evidence from mutual fund investment outcomes in China. Pacific-Basin Finance Journal. 2020;59. doi: 10.1016/j.pacfin.2019.101242.
11. Mohammadi A, Shafi K. Gender differences in the contribution patterns of equity-crowdfunding investors. Small Business Economics. 2018;50(2):275-87. doi: 10.1007/s11187-016-9825-7.
12. Cao Q, Niu X, Li J. Are women better traders?An experimental study. South China Journal of Economics. 2021;(07):128-44. doi: 10.19592/j.cnki.scje.380154.
13. Zhang Z, Zhang Y, Shen D, Zhang W. The influence of the mass media and new media on the return volatility in Chinese stock market. Chinese Journal of Management Science. 2021;29(06):238-48. doi: 10.16381/j.cnki.issn1003-207x.2018.1769.
14. Jin D, Li Y. Wisdom of crowds: Peer opinions and value discovery——Empirical evidences of social media. Business and Management Journal. 2017;39(12):157-73. doi: 10.19616/j.cnki.bmj.2017.12.010.
15. Jiang Y, Yan X. An information structure perspective: The information spillovers between Internet forum and stock market in China. South China Journal of Economics. 2015;(11):36-52. doi: 10.19592/j.cnki.scje.2015.11.003.
16. Zhao X, Wang Y. The empirical study on disposition effect in China's stock market. Journal of Financial Research. 2001;(07):92-7.
17. Da Costa N, Jr., Goulart M, Cupertino C, Macedo J, Jr., Da Silva S. The disposition effect and investor experience. Journal of Banking & Finance. 2013;37(5):1669-75. doi: 10.1016/j.jbankfin.2012.12.007. PubMed PMID: WOS:000317167600028.
18. da Silva PP, Mendes V. Exchange-traded certificates, education and the disposition effect. Journal of Behavioral and Experimental Finance. 2021;29. doi: 10.1016/j.jbef.2020.100456. PubMed PMID: WOS:000632618200019.
19. Hwang Y, Park J, Kim JH, Lee Y, Fabozzi FJ. Heterogeneous trading behaviors of individual investors: A deep clustering approach. Finance Research Letters. 2024;65. doi: 10.1016/j.frl.2024.105481. PubMed PMID: WOS:001244204900001.
20. Ben-David I, Hirshleifer D. Are investors really reluctant to realize their losses? Trading responses to past returns and the disposition effect. The Review of Financial Studies. 2012;25(8):2485-532. doi: 10.1093/rfs/hhs077. PubMed PMID: WOS:000306645600005.
21. Ülkü N, Ali F, Saydumarov S, İkizlerli D. COVID caused a negative bubble. Who profited? Who lost? How stock markets changed? Pacific-Basin Finance Journal. 2023;79. doi: 10.1016/j.pacfin.2023.102044.

22. Welch I. The wisdom of the Robinhood crowd. Journal of Finance. 2022;77(3):1489-527. doi: 10.1111/jofi.13128. PubMed PMID: WOS:000782476500001.
23. An L, Engelberg J, Henriksson M, Wang B, Williams J. The portfolio-driven disposition effect. Journal of Finance. 2024;79(5). doi: 10.1111/jofi.13378. PubMed PMID: WOS:001294870700001.
24. Xiong X, Luo C, Zhang Y. Stock BBS and trades: The information content of stock BBS. Journal of Systems Science and Mathematical Sciences. 2017;37(12):2359-74.
25. Zheng Y, Dong D, Zhu H. Does Internet communication of stock information weaken the stock market herding?An analysis of Chinese market. Management Review. 2015;27(06):58-67. doi: 10.14120/j.cnki.cn11-5057/f.2015.06.006.
26. Zheng J, Lv X, Lv B, Guo F. Information interaction on social media platforms and capital market pricing efficiency: Evidence from big data of stock message board. Journal of Quantitative & Technological Economics. 2022;39(11):91-112. doi: 10.13653/j.cnki.jqte.2022.11.005.
27. Xiao Z, Xie C, Chen Y. Can the noise comment in stock network platform affect the stock price synchronicity. Economic Theory and Business Management. 2021;41(10):65-80.
28. Sun Y, Cheng Q, Jin Q, Guo K. The impact of social interaction on the disposition effect of individual investors: Empirical analysis based on social trading platform. Management Review. 2020;32(10):72-82. doi: 10.14120/j.cnki.cn11-5057/f.2020.10.006.
29. Gemayel R, Preda A. Does a scopic regime erode the disposition effect? Evidence from a social trading platform. Journal of Economic Behavior & Organization. 2018;154:175-90. doi: 10.1016/j.jebo.2018.08.014. PubMed PMID: WOS:000454969500010.
30. Heimer RZ. Peer pressure: Social interaction and the disposition effect. The Review of Financial Studies. 2016;29(11):3177-209. doi: 10.1093/rfs/hhw063. PubMed PMID: WOS:000388026900009.
31. Frino A, Lepone G, Wright D. Investor characteristics and the disposition effect. Pacific-Basin Finance Journal. 2015;31:1-12. doi: 10.1016/j.pacfin.2014.10.009. PubMed PMID: WOS:000353855100001.
32. Fan P, Yang Y, Zhang Z, Chen M. The relationship between individual stock investor sentiment and the stock yield——Based on the perspective of stock evaluation information. Mathematics in Practice and Theory. 2021;51(16):305-20.
33. Cai Y, Tang Z, Wu J, Zhang T, Du X, Chen K. Research on the influence of heterogeneous investor emotion on stock market: Based on text semantic analysis. Journal of Systems Science and Mathematical Sciences. 2021;41(11):3093-108.
34. Fan X, Wang Y, Wang D, Guo W, Hu X. Heterogeneity analysis of information content for financial text from different sources: A hybrid text sentiment measurement method. Journal of Management World. 2022;38(10):78-101. doi: 10.19744/j.cnki.11-1235/f.2022.0145.
35. Jiang F, Meng L, Tang G. Media textual sentiment and Chinese stock return predictability. China Economic Quarterly. 2021;21(04):1323-44. doi: 10.13821/j.cnki.ceq.2021.04.10.
36. Feng L, Seasholes MS. Do investor sophistication and trading experience eliminate behavioral biases in financial markets? Review of Finance. 2005;(NO.3):305-51.
37. Seru A, Shumway T, Stoffman N. Learning by trading. The Review of Financial Studies. 2010;(No.2):705-39.
38. Tan S, Chen Y. Do individual investors learn from their trading experience? Journal of Financial Research. 2012;(05):164-78.
39. Zhang K, Li X, Fang X, Li X. Opinion leader at the online stock forum: Having the discerning eye or eye-catching. Journal of Management Sciences in China. 2022;25(09):90-107. doi:

10.19920/j.cnki.jmsc.2022.09.006.